\documentstyle[draft]{laa}

\def\lapp{\ifmmode\stackrel{<}{_{\sim}}\else$\stackrel{<}{_{\sim}}$\fi}
\def\gapp{\ifmmode\stackrel{>}{_{\sim}}\else$\stackrel{>}{_{\sim}}$\fi}
\newcommand{\PSRA}{J0215$+$6218}
\newcommand{\PSRB}{J1957$+$2831}
\newcommand{\SNRA}{G132.7$+$1.3}
\newcommand{\SNRB}{G65.1$+$0.6}

\begin{document}

\title{A search for pulsars in supernova remnants}

\author{ D.R.~Lorimer\inst{1}, A.G.~Lyne\inst{2}, F.~Camilo\inst{2}}

 \offprints{ D.R.~Lorimer (E-mail: dunc@mpifr-bonn.mpg.de)}
 \institute{ Max-Planck-Institut f\"ur Radioastronomie, Auf dem H\"ugel 69,
             D-53121 Bonn, Germany 
 \and        University of Manchester, Nuffield Radio Astronomy Laboratories,
             Jodrell Bank, Macclesfield, Cheshire, SK11 9DL, UK
              }

\thesaurus{ 08.16.7 ; PSR {\PSRA} ; PSR {\PSRB} ; 09.19.2 ; {\SNRA}; {\SNRB} }

\maketitle
\markboth{Lorimer et al.: A search for pulsars in supernova remnants}{}

\begin{abstract} 

We have carried out a sensitive search for young pulsars associated
with supernova remnants using the 76--m Lovell radio telescope at
Jodrell Bank. The observations were made at 606 MHz using a system
with a bandwidth of 8 MHz and a set noise temperature on cold sky of
about 50 K.  The survey targeted 33 remnants in the northern
hemisphere and achieved a nominal sensitivity of $\sim 1$ mJy in most
cases. Two pulsars were discovered in the course of this survey and
the known pulsar PSR~B1952+29 was detected. The new pulsars, {\PSRA}
and {\PSRB}, were found during searches of the supernova remnants
{\SNRA} and {\SNRB} respectively. Based on a statistical analysis of
the present sample of proposed pulsar--supernova remnant pairs, we
conclude that at most 17 associations are likely to be real. We find
no strong evidence for a genuine association between either of the two
newly discovered pulsars and their target supernova remnants.


\end{abstract}

\section{Introduction}
\label{sec:intro}

In the standard model for pulsar evolution (Gunn \& Ostriker 1970),
\nocite{go70} young rapidly rotating pulsars are expected to be
harboured within the remnants of the supernova explosions in which
they were formed. The two best--known examples of young pulsars, {\it
viz.} Crab and Vela, were discovered in the early days of pulsar
astronomy (Staelin \& Reifenstein 1968; Large et
al.~1968). \nocite{sr68,lvm68} The overwhelming evidence in favour of
their association with the Crab and Vela supernova remnants helped to
establish the basic rotating neutron star model to explain the pulsar
phenomenon (Pacini 1968; Gold 1968). \nocite{pac68,gol68} In this
paper, we describe a survey carried out to find pulsars associated
with supernova remnants in the northern hemisphere.

In the past few years, the number of {\it proposed associations}
between pulsars and supernova remnants has risen to around 30 (see
\nocite{kas96} Kaspi 1996 for a recent review). This is a result of a
number of different approaches: high frequency searches of the
Galactic plane (Clifton et al.~1992; Johnston et al.~1992);
\nocite{clj+92,jlm+92} cross--correlations of the pulsar and supernova
remnant catalogues \nocite{car93,frs93} (Caraveo 1993; F\"urst et
al.~1993; Kulkarni et al.~1993); \nocite{kpha93} targeted searches for
pulsars in remnants (Frail \& Moffett 1993; Gorham et al.~1996; Kaspi
et al.~1996) \nocite{fm93,gra+96,kmj+96} and remnants around young
pulsars (Frail et al.~1994).  \nocite{fgw94}

The number of spurious associations in this sample, {\it i.e.} those
cases where the pulsar and the supernova remnant are merely in chance
alignment on the sky, is likely to be significant.  Theoretically, the
expected number of associations has a complex dependence on a number
of factors: the luminosity function and initial spin period of young
pulsars (Narayan 1987; Narayan \& Schaudt 1988), \nocite{nar87,ns88}
their beaming fraction (Lyne \& Manchester 1988; Biggs 1990),
\nocite{lm88,big90b} space velocities (Shull et al.~1989; Caraveo
1993; Lyne \& Lorimer 1994), \nocite{sfs89,car93,ll94} the evolution
of supernova remnants (Gaensler \& Johnston 1995a\&b)
\nocite{gj95a,gj95c} as well as the fraction of all supernovae that
produce pulsars.  In a detailed statistical analysis, Gaensler \&
Johnston (1995b) \nocite{gj95c} concluded that fewer than 10
associations are expected to be real.

Statistical studies require accurate flux density upper limits for any
pulsars undetected in remnants.  Our search complements recent
targeted searches on southern supernova remnants at Parkes (Kaspi et
al.~1996), \nocite{kmj+96} on those remnants visible from Arecibo
(Gorham et al.~1996), \nocite{gra+96} as well as an earlier survey at
Jodrell Bank (Biggs \& Lyne 1996). \nocite{bl96} The present survey
was successful, discovering two new pulsars as well as substantially
improving on previous upper limits for the flux density of any pulsars
in the target remnants.

\begin{table*}
\begin{center}
\begin{tabular}{llllllrrrl}
\hline
Remnant & Alias & $\alpha_{1950}$ & $\delta_{1950}$&size&Type & $T_{\rm sky}$ &
 $T_{\rm rem}$ &  $N_{\rm pnt}$ & $S_{\rm min}$\\
   Name   &       &h \,\,\,m \,\,\,s  &   deg \,\,m  & arcmin &    &   K  & K   
             &                  &  mJy  \\
\hline
G65.1+0.6  &              &  19 52 30  &  +28 25 &  90x50 &S &30&1 &7 &\hspace{.17cm}0.8\\
G65.7+1.2  & DA 495       &  19 50 10  &  +29 18 &  18    &? &30&7 &1 &\hspace{.17cm}0.9\\
G67.7+1.8  &              &  19 52 34  &  +31 21 &  9     &S &30&2 &1 &\hspace{.17cm}0.8\\
G68.6$-$1.2  &            &  20 06 40  &  +30 28 &  28x25?&? &31&1 &1 &\hspace{.17cm}0.9\\
G69.7+1.0  &              &  20 00 45  &  +32 35 &  16    &S &34&2 &1 &\hspace{.17cm}0.9\\
& & & & & & & & \\
G73.9+0.9  &              &  20 12 20  &  +36 03 &  22?   &S?&42&10&1 &\hspace{.17cm}1.1\\
G74.9+1.2  & CTB 87       &  20 14 10  &  +37 03 &  8x6   &F &43&11&1 &\hspace{.17cm}1.1\\
G76.9+1.0  &              &  20 20 30  &  +38 33 &  9x12  &? &77&3 &1 &\hspace{.17cm}1.3\\
G78.2+2.1  &$\gamma-$Cygni&  20 19 00  &  +40 15 &  60    &S &72&110&7&\hspace{.17cm}2.4\\
G82.2+5.3  & W63          &  20 17 30  &  +45 20 &  95x65 &S &33&15&7 &\hspace{.17cm}1.0\\
& & & & & & & & \\
G84.2$-$0.8  &            &  20 51 30  &  +43 16 &  20x16 &S &44&14&1 &\hspace{.17cm}1.1\\
G84.9+0.5  &              &  20 48 45  &  +44 42 &  6     &S &28&1&1 &\hspace{.17cm}0.8\\
G89.0+4.7  & HB21         &  20 43 30  &  +50 25 &  120x90&S &26&17&13&\hspace{.17cm}1.0\\
G93.3+6.9  & DA530        &  20 51 00  &  +55 10 &  27x20 &S &18&12&1 &\hspace{.17cm}0.8\\
G93.7$-$0.2  & CTB104A    &  21 27 45  &  +50 35 &  80    &S &25&11&7 &\hspace{.17cm}0.9\\
& & & & & & & & \\
G94.0+1.0  & 3C434.1      &  21 23 10  &  +51 40 &  30x25 &S &26&19&1 &\hspace{.17cm}1.0\\
G109.1$-$1.0 & CTB109     &  22 59 30  &  +58 37 &  28    &S&150&26&1 &\hspace{.17cm}2.3\\
G111.7$-$2.1 & Cass--A    &  23 21 10  &  +58 32 &  5     &S&300&4000&1&46\\
G112.0+1.2 &              &  23 13 40  &  +61 30 &  30?   &S?&31&9&1 &\hspace{.17cm}0.9\\
G116.5+1.1 &              &  23 51 20  &  +62 58 &  80x60 &S &25&2&7 &\hspace{.17cm}0.8\\
& & & & & & & & \\
G116.9+0.2 & CTB1         &  23 56 40  &  +62 10 &  34    &S &22&9&1 &\hspace{.17cm}0.8\\
G117.4+5.0 &              &  23 52 30  &  +67 30 &  60x80?&S?&22&5&7 &\hspace{.17cm}0.8\\
G119.5+10.2& CTA1         &  00 04 00  &  +72 30 &  90?   &S &17&5&7 &\hspace{.17cm}0.8\\
G120.1+1.4 & Tycho SN1572 &  00 22 30  &  +63 52 &  8     &S &25&76&1 &\hspace{.17cm}1.6\\
G126.2+1.6 &              &  01 18 30  &  +64 00 &  70    &S?&23&2&7 &\hspace{.17cm}0.8\\
& & & & & & & & \\
G127.1+0.5 & R5           &  01 25 00  &  +62 55 &  45    &S &23&8&1 &\hspace{.17cm}0.8\\
G130.7+3.1 & 3C58 SN1181  &  02 01 55  &  +64 35 &  9x5   &F &21&35&1 &\hspace{.17cm}1.1\\
G132.7+1.3 & HB3          &  02 14 00  &  +62 30 &  80    &S &27&7&7 &\hspace{.17cm}0.9\\
G152.2$-$1.2 &            &  04 05 30  &  +48 24 &  110?  &S?&21&2&7 &\hspace{.17cm}0.8\\
G156.2+5.7 &              &  04 54 40  &  +51 47 & 110    &S &16&1&7 &\hspace{.17cm}0.7\\
& & & & & & & & \\
G166.0+4.3 & VRO 42.05.01 &  05 23 00  &  +42 52 & 55x35  &S &14&3&7 &\hspace{.17cm}0.7\\
G166.2+2.5 & OA 184       &  05 15 30  &  +41 50 &  90x70 &S &14&2&7 &\hspace{.17cm}0.7\\
G179.0+2.6 &              &  05 50 30  &  +31 05 &  70    &S?&12&2&7 &\hspace{.17cm}0.7\\
\hline
\end{tabular}
\end{center}
\caption[]
{
33 supernova remnants targeted by the survey. From left to right the
columns give the remnant name based on its Galactic coordinates, any
alias(es) by which the remnant may be called, the equatorial
coordinates, the approximate angular size, a classification tag (S --
shell remnant, ? -- unknown type, F -- filled centre).  Question marks
denote poorly-known observational quantities. All these data are taken
from Green's (1996) catalogue of supernova remnants \nocite{gre96b} to
which the interested reader is referred for further details. We also
list the estimated contribution to the system temperature from the sky
background ($T_{\rm sky}$) and the remnant itself ($T_{\rm rem}$), the
number of telescope pointings used to observe each remnant ($N_{\rm
pnt}$), as well as the estimated minimum flux density at 606~MHz
($S_{\rm min}$), above which a pulsar would have been detected by the
survey (see text).
}
\label{tab:snrs}
\end{table*}

The plan for this paper is as follows: In Sect.~\ref{sec:obs} we
describe the survey observations and data reduction techniques. In
Sect.~\ref{sec:sens} we estimate the sensitivity of the survey. We
summarise the results of the survey in Sect.~\ref{sec:res}.  In
Sect.~\ref{sec:disc} we discuss the validity of the association
between the detected pulsars and their target remnants and comment on
the implications of the flux density upper limits for the luminosity
of pulsars at birth. In Sect.~\ref{sec:stats} we carry out a simple
model-free statistical analysis to deduce the number of associations
that would be expected by chance.  Finally in Sect.~\ref{sec:conc} we
present our conclusions.

\section{Survey Observations} 
\label{sec:obs}

In order to complement recent searches in the southern hemisphere and
from Arecibo, we chose to search primarily those supernova remnants
with declinations north of 30 degrees.  Our final sample of 33
supernova remnants was selected from Green's (1996) catalogue and is
summarised in Table \ref{tab:snrs}.

The survey observations were made with the 76--m Lovell radio
telescope operated by the University of Manchester at Jodrell Bank, on
5 separate observing sessions between 1994 June and 1996 April. The
centre frequency for the observations was 606 MHz with a total
bandwidth of 8 MHz. For those remnants larger than the telescope beam
width (0.5 degrees FWHM) we covered the full area of the remnant using
up to 13 individual telescope pointings (see Table \ref{tab:snrs}).
Each telescope pointing was of 35 minutes duration.

The incoming radiation was split by the feed into two orthogonal
linear polarisations and amplified by a pair of cryogenically-cooled
field effect transistors. The signals then entered wide-band filters
and, following a second stage of amplification, were mixed down to an
intermediate frequency centred on 8 MHz before entering a pair of $64
\times 0.125$ MHz filterbanks. The resulting 64 polarisation pairs
were then detected, summed, integrated and one-bit digitised every
millisecond.  The data were then passed to an on-line VAX computer
which wrote the data to a standard Exabyte magnetic tape for off-line
processing.

The off-line search for periodic signals in the data consisted of a
Fourier analysis of time sequences of dedispersed data in order to
search for significant features in the power spectrum.  This was
followed by a time series analysis to optimise the parameters of the
most promising spectral features. The complete search procedure is
very similar to that used for the highly successful Parkes southern
sky survey described by \nocite{mld+96} Manchester et al.~(1996).
Briefly, the data were de-dispersed and Fourier transformed for 90
trial values of dispersion measure (DM) between zero and 995 cm$^{-3}$
pc. For any periodic signal with a small duty cycle (5--10\% is
typical of most pulsars), the resulting power spectrum from the
Fourier transform consists of a family of harmonic spikes with the
fundamental corresponding to the signal frequency.  To increase the
sensitivity to pulsars with narrow pulses, higher order harmonics can
be added onto the fundamental (Lyne 1988)\nocite{lyn88}.  For example,
to add all 2$^{\rm nd}$ harmonics onto their corresponding
fundamentals, we stretch the lower half of the amplitude spectrum by a
factor of two and add this to the original unstretched spectrum. While
the effect of this harmonic summing process increases the noise by a
factor of $\sqrt{2}$, the amplitude of the signal may increase by more
than a factor of $\sqrt{2}$ giving a net increase in the
signal-to-noise ratio. By repeating this process several times we
effect a search in pulse duty cycle. In our analysis the harmonic
summing was performed 4 successive times, {\it i.e.}  producing power
spectra summed over 1, 2, 4, 8 and 16 harmonics. During this stage
periodic signals from known sources of terrestrial interference, {\it
e.g.} the 50 Hz mains power line, were excised from the data by
zeroing the appropriate portions of the power spectrum. Due to strong
amounts of terrestrial interference in the low-frequency part of the
power spectrum, we did not consider signals with frequencies below 0.2
Hz.  This degraded our sensitivity only slightly to pulsars with
periods greater than 5 seconds which could still be detected in their
higher order harmonics.

For each value of DM, the largest features in the power spectra and
harmonic sums were then sorted so as to produce a list of ``pulsar
suspects''. All suspects with spectral signal-to-noise ratios greater
than 7 were then folded in the time domain for a range of periods and
DMs around the nominal values found in the first stage in an attempt
to optimise these parameters and the signal-to-noise ratio. The output
from this stage was visually inspected (see also Fig.~5 of Manchester
et al.~1996) and the best suspects were stored for re-observation.

\section{Search Sensitivity}
\label{sec:sens}

The sensitivity of the survey can be expressed in terms of the minimum
detectable flux density $S_{\rm min}$ which is a function of a number
of parameters. Following Dewey et al.~(1984), we write: \nocite{dss+84}
\begin{equation}
\label{equ:smin}
   S_{\rm min} = \frac{\sigma \eta \, T_{\rm sys}}{G \sqrt{n \Delta
        \nu \tau}} \left(\frac{W}{P-W} \right)^{1/2}.
\end{equation}
In this expression $\sigma$ is the threshold signal-to-noise ratio
above which a detection is considered significant (7 in our case),
$\eta$ is a constant $\sim 1.3$ which reflects losses to hardware
limitations, $G$ is the gain of the telescope (1.0 K Jy$^{-1}$ for the
Lovell telescope operating at 606 MHz), $n$ is the number of
polarisations used, $\Delta \nu$ is the observing bandwidth, $T_{\rm
sys}$ is the system temperature, $\tau$ is the integration time, $P$
is the period of the pulsar and $W$ is the observed width of the
pulse. With the parameters for this survey, the above expression
simplifies to
\begin{equation}
\label{equ:smin2}
S_{\rm min} \simeq 0.05 \, T_{\rm sys}
\left(\frac{W}{P-W}\right)^{1/2} \, {\rm mJy}.
\end{equation}

The system temperature $T_{\rm sys}$ is the sum of separate
components: the set noise of the receiver $T_{\rm set}$; the sky
background noise $T_{\rm sky}$ and the contribution $T_{\rm rem}$ from
continuum flux of the supernova remnant. Regular calibration
measurements made during the survey indicated $T_{\rm set}$ to be
typically 50 K. The contribution to $T_{\rm sys}$ from the Galactic
background and the supernova remnant covers a large range.  This is
shown in Table \ref{tab:snrs} where we list the estimated sky
background temperatures from a machine-readable version of the Haslam
et al.~(1982) \nocite{hssw82} all-sky survey, scaled to 606 MHz
assuming a spectral index of --2.7 (Lawson et al.~1987)
\nocite{lmop87} together with the expected contribution from each
supernova remnant. To calculate the latter values, we estimated the
flux density of the remnant at 606 MHz from spectral information in
Green's (1996) catalogue and multiplied this by the beam filling
factor, defined as the lesser of unity and $\left({\rm FWHM}/D_{\rm
SNR}\right)^2$, where $D_{\rm SNR}$ is the angular diameter of the
supernova remnant.  Using these values in equation \ref{equ:smin2}, we
have estimated the minimum flux density required to detect a 0.1 s
pulsar with a duty cycle of 4\% in each remnant. These limiting flux
densities are listed in Table \ref{tab:snrs} for reference. With
typical values of $\sim$ 1 mJy at 606 MHz, they demonstrate the
excellent sensitivity of the survey. We note from Table \ref{tab:snrs}
that $T_{\rm rem}$ is a significant factor for only 3 of the 33 SNRs
searched.

The observed pulse width $W$ in Eqs.~\ref{equ:smin} and
\ref{equ:smin2} is likely to be greater than the intrinsic width
$W_{\rm int}$ emitted at the pulsar because of the scattering and
dispersion of pulses by free electrons in the interstellar medium, and
by the post-detection integration performed in the receiver.  The
observed sampled pulse profile will therefore be the convolution of
the intrinsic pulse width and broadening functions due to dispersion,
scattering and integration and can be estimated approximately from the
following quadrature sum:
\begin{equation}
  W^2 = W_{\rm int}^2 + t_{\rm samp}^2 + t_{\rm DM}^2 + t_{\rm scatt}^2,
\end{equation} 
where $t_{\rm samp}$ is the data sampling interval, $t_{\rm DM}$ is
the dispersion broadening across one filterbank channel and $t_{\rm
scatt}$ is the interstellar scatter broadening. Pulse scattering
becomes particularly important when observing distant pulsars towards
the inner Galaxy at frequencies $\lapp$ 1 GHz. Many of the supernova
remnants in our sample are relatively nearby $\lapp$ 5 kpc so that we
do not expect a significant effect on our sensitivity due to
scattering.

\begin{figure}
\setlength{\unitlength}{1in}
\begin{picture}(0,2.6)
\put(-0.3,2.7){\includegraphics{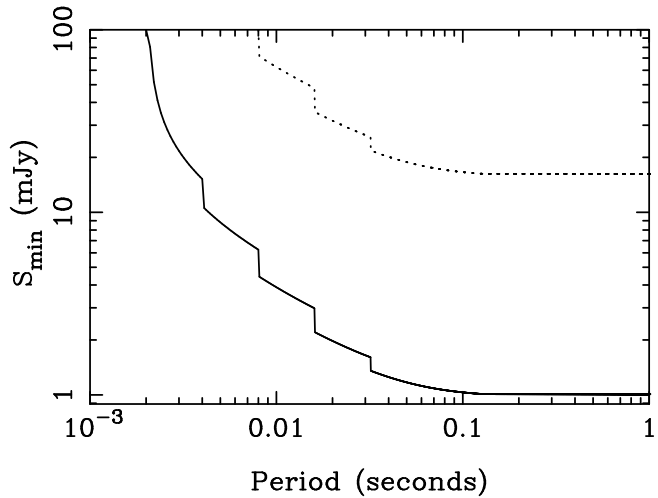}}
\end{picture}
\caption[]
{
The limiting sensitivity of the survey as a function of pulse period
for a pulsar with a DM of 150 cm$^{-3}$\,pc and a duty cycle of 4\% is
shown by the solid curve.  The calculation assumes a total system
temperature of $\sim$ 95 K, a typical value for supernova remnants
observed during the survey. The dotted curve shows the approximate
sensitivity to a pulsar lying 0.25 degrees outside the boundary of the
remnant (see text).
}
\label{fig:smin}
\end{figure}

The effects of sampling and dispersion do, however, significantly
affect the search sensitivity at short pulse periods. This is shown in
Fig.~\ref{fig:smin}, where $S_{\rm min}$ is plotted against $P$ for a
pulsar with a dispersion measure of 150 cm$^{-3}$\,pc, a typical value
for a pulsar at the distance of one of the remnants in our sample (see
Sect.~\ref{ssec:lum}). We have also calculated the reduction in
sensitivity at short periods due to the loss of higher order harmonics
in the power spectrum -- shown by the abrupt jumps in the sensitivity
curve shown in Fig.~\ref{fig:smin}.

It is worth noting that our search had only limited sensitivity to any
high-velocity pulsars that may have moved outside the projected
boundaries of their parent supernova remnants. The dotted curve in
Fig.~\ref{fig:smin} gives the approximate sensitivity to a pulsar
lying 0.25 degrees outside the remnant boundary and indicates a
limiting 600 MHz flux density $\gapp 20$ mJy. Such bright pulsars should
have been found in previous large scale surveys of the northern
sky. Future surveys targeted specifically outside the periphery of
the remnants will substantially improve the sensitivity to fainter
pulsars.

\vspace{1cm}

\section{Results and follow-up observations}
\label{sec:res}

Twenty-five of the most promising pulsar candidates were re-observed
in April 1996 and the detection of three pulsars was confirmed,
B1952+29 and two that were previously unknown: PSRs {\PSRA} and
{\PSRB}. No further previously known pulsars were expected to be above
our detection threshold.  The new pulsars lie within the boundaries
but towards the outer regions of the target supernova remnants {\SNRA}
and {\SNRB} respectively. No further pulsars were discovered during
these observations.

In order to accurately determine the astrometric and spin parameters
of the two newly discovered pulsars, regular follow-up observations at
606 and 1400 MHz have been made ever since their discovery as part of
the Jodrell Bank pulsar timing programme (see for example Shemar \&
Lyne 1996). \nocite{sl96} We have been able to model the pulse arrival
times for both pulsars using standard pulsar timing techniques
(Manchester \& Taylor 1977). \nocite{mt77} Timing solutions obtained
from these data are given in Table \ref{tab:psrs}.

\begin{table}
\begin{center}
\begin{tabular}{lll}
\hline
PSR              &      {\PSRA}     &     {\PSRB}     \\
\hline
R. A. (J2000)    & 02 15 56.61(1)   & 19 57 19.36(3)  \\
Decl. (J2000)    & 62 18 33.36(6)   & 28 31 44.1(2)   \\
Period (sec)     & 0.548879759260(4)& 0.30768255585(6)\\
$\dot{P} (10^{-15})$& 0.661(6)& 3.124(7)       \\
Epoch (MJD)      & 50295.0          & 50278.0         \\
DM    (cm$^{-3}$\,pc) & 84.22(2)    & 139.08(2)       \\
$S_{600}$ (mJy)  & 10.6(4)          & 2.1(5)          \\
$S_{1400}$ (mJy) & 3.7(7)           & 1.0(2)          \\
\hline
$\alpha$         & $-$1.3(2)        & $-$0.9(4)        \\
$\tau_c$ (Myr)   & 13               & 1.6             \\
$B$ ($10^{12}$ G) & 0.61     & 0.99            \\
\hline
\end{tabular}
\end{center}
\caption[]
{
Observed and derived parameters of the two newly discovered pulsars.
Quoted errors refer to the last significant digit.
}
\label{tab:psrs}
\end{table}

\begin{figure}
\setlength{\unitlength}{1in}
\begin{picture}(0,2.4)
\put(-0.4,2.8){\includegraphics{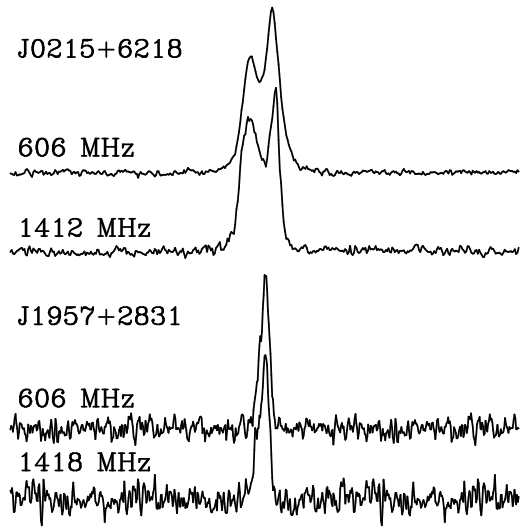}}
\end{picture}
\caption[]
{
Total intensity integrated pulse profiles for PSRs {\PSRA} and
{\PSRB}. In each case the whole of the pulsar period is displayed.
These profiles are freely available at the on-line database of pulse
profiles --- http://www.mpifr-bonn.mpg.de/pulsar/data.
}
\label{fig:profs}
\end{figure}

Total intensity pulse profiles of the newly discovered pulsars at 606
and 1412/1418 MHz are shown in Fig.~\ref{fig:profs}. Mean flux
densities at both frequencies were determined in an identical fashion
to that described by Lorimer et al.~(1995).  \nocite{lylg95} We also
list the characteristic age $\tau_c = P/2\dot{P}$ and the surface
magnetic field $B \approx 3.2 \times 10^{19} (P\dot{P})^{1/2}$ Gauss
which assume a constant dipolar magnetic field and short initial
period (Manchester \& Taylor 1977). \nocite{mt77}

\section{Discussion}
\label{sec:disc}

\subsection{Are the detected pulsars really associated with their 
target supernova remnants?}
\label{ssec:assoc}

The previously known pulsar PSR B1952+29 was detected during
observations of G65.1+0.6. The pulsar lies outside the boundary of the
remnant and is clearly not associated with G65.1+0.6 because of its
extremely large characteristic age ($4 \times 10^{9}$ yr) and the
proper motion measurement (Lyne et al.~1982), \nocite{las82} which
shows that it is moving towards rather than away from the remnant
centre as is required for a genuine association.

The evidence that the newly discovered pulsars are associated with
their target remnants is not clear. PSR {\PSRA} has a period of 549 ms
and a DM of 84 cm$^{-3}$ pc. The distance to this pulsar estimated
from its DM and Galactic coordinates using the Taylor \& Cordes (1993)
\nocite{tc93} electron density model is $3.2\pm1.0$ kpc, consistent
with the distance to {\SNRA} of $2.2 \pm 0.2$ kpc \nocite{rdlv91}
(Routledge et al.~1991). Timing measurements show that the
characteristic age of PSR {\PSRA} is 13 Myr, anomalously large by
comparison with the \nocite{lvln85} age of {\SNRA}, estimated to be
between 30,000 and 50,000 yr (Leahy et al.~1985).

In the case of the 308 ms pulsar {\PSRB}, there is unfortunately no
independent distance estimate to the target remnant, {\SNRB}. However
the angular size of the remnant ($\sim 1$ degree) indicates that it is
likely to be closer than the distance of $7.0\pm2.3$ kpc inferred from
the dispersion measure of PSR {\PSRB} (139 cm$^{-3}$ pc). Whilst the
age of {\SNRB} is also not known, it is likely to be much smaller than
the characteristic age of PSR {\PSRB} (1.6 Myr).

Thus, in both cases, the pulsar characteristic age is anomalously
large by comparison with the expected ages of the supernova remnants.
This suggests that they are either not associated with the remnant, or
the characteristic ages are anomalously large, due perhaps to the
initial spin period of the pulsar being similar to its presently
observed value.  We note in passing that the spectral index of PSRs
{\PSRA} and {\PSRB} are --1.2 and --0.9 respectively, typical of many
other young pulsars (Lorimer et al.~1995) \nocite{lylg95} and
suggesting that these pulsars may indeed be younger than their
characteristic ages would suggest.  Although this remains a
possibility, as we shall show in Sect.~\ref{sec:stats}, on statistical
grounds, neither of the newly discovered pulsars are likely to be
associated with the target supernova remnants.

\subsection{The luminosity of pulsars at birth}
\label{ssec:lum}

In a study of pulsar population statistics Lorimer et al.~(1993)
\nocite{lbdh93} suggested that there may be no need for a significant
number of pulsars to be born with 400 MHz radio luminosities below 30
mJy kpc$^2$.  Deep surveys of supernova remnants, like the present
one, can in principle be used to test this hypothesis by combining the
flux density upper limit ($S_{\rm min}$) given in Table \ref{tab:snrs}
with the distance to each remnant ($d_{\rm SNR}$) to estimate the
minimum luminosity ($L_{\rm min}$) that a young pulsar would need in
order to be detectable.  In Table \ref{tab:dist} we list the 17
supernova remnants in our sample which have reliable distance
estimates.  Note that we have chosen not to use the surface
brightness--angular size ($\Sigma-D$) relationship (Clark \& Caswell
1976) \nocite{cc76} as a means of estimating $d_{\rm SNR}$ since it
has subsequently been shown to be unreliable \nocite{ber87,gre91}
(Berkhuijsen 1987; Green 1991).

\begin{table}
\begin{center}
\begin{tabular}{llrlr}
\hline
  Remnant & Alias & Dist. & Ref.          &$L_{\rm min}$\\
   Name   &       &  kpc     &            & mJy kpc$^2$ \\
\hline
 G73.9+0.9  &              & 1.3 &  a     &  4          \\
 G74.9+1.2  & CTB 87       & 12 &  b     &  320        \\
 G78.2+2.1  &$\gamma-$Cygni& 1.5 &  c,d   &  11         \\
 G84.2$-$0.8  &            & 4.5 &  e     &  44         \\
 G89.0+4.7  & HB21         & 0.8 &  f     &  1          \\
& & & & \\
 G109.1$-$1.0 & CTB109     & 4.0 &  d     &  74         \\
 G111.7$-$2.1 & Cass--A    & 2.8 &  g     &  720        \\
 G116.5+1.1 &              & 4.4 &  h     &  30         \\
 G116.9+0.2 & CTB1         & 3.1 &  i     &  15         \\
 G119.5+10.2& CTA1         & 1.4 &  j     &  3          \\
& & & & \\
 G120.1+1.4 & Tycho SN1572 & 2.7 &  k     &  22         \\
 G127.1+0.5 & R5           & 1.3 &  l     &  3          \\
 G130.7+3.1 & 3C58 SN1181  & 3.2 &  m     &  22         \\
 G132.7+1.3 & HB3          & 2.2 &  n     &  9          \\
 G156.2+5.7 &              & 2.0 &  o     &  6          \\
& & & & \\
 G166.0+4.3 & VRO 42.05.01 & 4.5 &  p     &  28         \\
 G166.2+2.5 & OA 184       & 8.0 &  q     &  90         \\
\hline
\end{tabular}
\end{center}
\caption[]
{
The supernova remnants targeted by the survey with published distance
estimates. From left to right the columns give the remnant name based
on its Galactic coordinates, alias(es) by which the remnant may be
called, the distance and reference tag, and the corresponding minimum
400 MHz luminosity for a pulsar to be detectable in our survey (see
text).  The references are: a. \nocite{lsp93} Lozinskaya et
al.~(1993); b.  \nocite{gg89} Green \& Gull (1989); c. \nocite{lrh80}
Landecker et al.~(1980); d. \nocite{gre89} Green (1989);
e. \nocite{fg93} Feldt \& Green (1993); f. \nocite{tflr90} Tatematsu
et al.~(1990); g. \nocite{van71} van den Bergh (1971);
h. \nocite{rb81b} Reich \& Braunsfurth (1981); i.  \nocite{hc94}
Hailey \& Craig (1994); j. \nocite{plmb93} Pineault et al.~(1993);
k. \nocite{atsg86} Albinson et al.~(1986); l. \nocite{gre96b} Green
(1996); m. \nocite{rgk+93} Roberts et al.~(1993); n. \nocite{rdlv91}
Routledge et al.~(1991); o. \nocite{rfa92} Reich et al.~(1992);
p. \nocite{lprv89} Landecker et al.~(1989); q. \nocite{rlv86}
Routledge et al.~(1986)
}
\label{tab:dist}
\end{table}

For each supernova remnant in Table 3, we list the minimum detectable
luminosity $L_{\rm min} = 2 S_{\rm min} d_{\rm SNR}^2$, where the
factor of 2 in this expression scales $S_{\rm min}$ defined for this
survey at 606 MHz to 400 MHz assuming a typical pulsar spectral index
of --1.6 (Lorimer et al.~1995) \nocite{lylg95}.  We find the median
value of $L_{\rm min}$ to be 22 mJy kpc$^2$, with 12 of these values
lying between 1 and 30 mJy kpc$^2$.  In this sample of 12 supernova
remnants with well defined distances, we detected one pulsar (PSR
{\PSRA}) which has a 400 MHz luminosity $\sim 80$ mJy kpc$^2$.  If we
assume that each of these remnants contains a radio pulsar born with a
luminosity above 30 mJy kpc$^2$, then we expect to see $12f$ pulsars,
where $f$ is the mean beaming fraction. A consensus on the beaming
fraction of young pulsars has yet to be established. Frail \& Moffet
(1993) \nocite{fm93} obtained a value $f \sim 0.6$ based on a deep
imaging search at the VLA. In this case we would expect to see $\sim
7$ pulsars. On the other hand, Tauris \& Manchester (1997)
\nocite{tm97} recently claim $f$ to be as low as $\sim 0.1$ implying a
detection of only one pulsar. Thus, although our results are
consistent with few pulsars being born with luminosities below 30 mJy
kpc$^2$, they are still hampered by small number statistics and we
conclude that yet deeper searches are required to provide a larger
sample with which to test this hypothesis (see also Kaspi et
al.~1996).

\subsection{Radio--quiet neutron stars}
\label{ssec:rquiet}

The above discussion does not, of course, exclude the possibility that
neutron stars exist in the remnants which do not produce significant
radio emission. One classic example is the 6.9 s X--ray pulsar 1E
2259+586 \nocite{fg81} which has long been postulated to be associated
with CTB 109 (Fahlman \& Gregory 1981).  Despite extensive efforts, no
radio counterpart to the pulsar has ever been detected. Coe et
al.~(1994) \nocite{cjl94} used the VLA at 1489 MHz to set a 3 $\sigma$
upper limit to continuum emission of 50 $\mu$Jy. Our limit to {\it
pulsed} emission from CTB 109 2.3 mJy at 600 MHz corresponds to about
500 $\mu$Jy when scaled to 1489 MHz, again assuming spectral index of
--1.6. Another example is 3C 58, a Crab-like plerion containing an
X-ray point source which is higly suggestive of a young pulsar
(Helfand et al.~1995 \nocite{hbw95} and references therein). Our limit
of 1.1 mJy to pulsed emission at 600 MHz corresponds, after scaling,
to about a factor of two larger than the present best upper limit of
0.15 mJy at 1400 MHz \nocite{fm93} (Frail \& Moffet 1993).

One of the supernova remnants observed during this survey, G78.2+2.1,
also known as $\gamma$--Cygni, contains the bright $\gamma$--ray
source 2EG J2020+4026. Brazier et al.~(1996) \nocite{bkc+96} have
recently discovered an X--ray source within the error box of 2EG
J2020+4026. Assuming this to be the X--ray counterpart, they propose,
on the basis of the $\gamma$--ray to X--ray flux ratio, that this
source is most probably a Geminga-like neutron star. No significant
radio pulsations were detected during our search of this remnant and
our estimated 400 MHz lower luminosity limit of 11 mJy kpc$^2$
certainly rules out the presence of a bright, favourably beamed,
Crab-like radio pulsar associated with this source of emission.

\begin{figure*}
\setlength{\unitlength}{1in}
\begin{picture}(0,3.0)
\put(1.0,3.5){\includegraphics{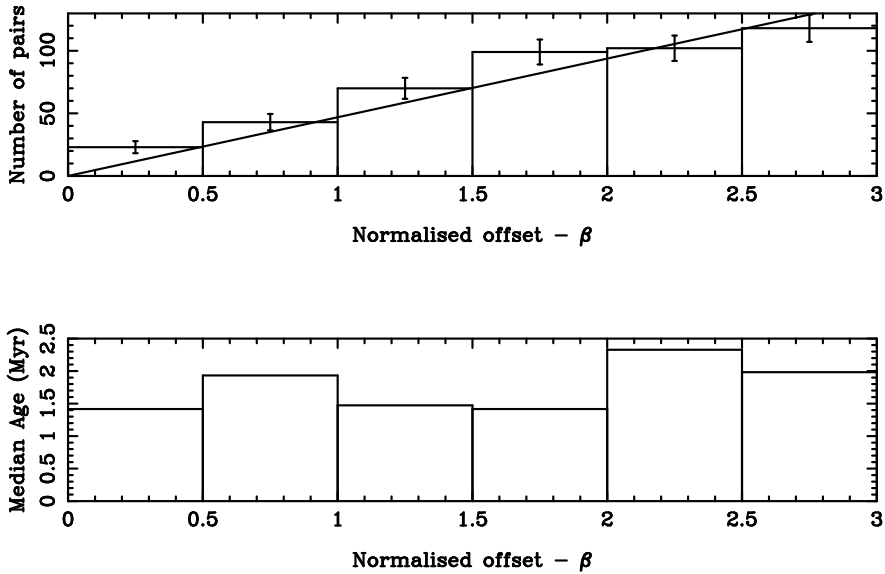}}
\end{picture}
\caption[]
{
Top panel: The distribution of $\beta$ for the sample of pulsar and
supernova remnants {\it after} applying shifts in Galactic longitude
to the pulsar sample; this is in excellent agreement with the
distribution $dN \propto \beta d \, \beta$ shown by the solid
line. The lower panel shows the median characteristic age of the
pulsars in the pairs as a function of $\beta$ (see text).
}
\label{fig:diddle}
\end{figure*}

\begin{figure*}
\setlength{\unitlength}{1in}
\begin{picture}(0,3.0)
\put(1.0,3.5){\includegraphics{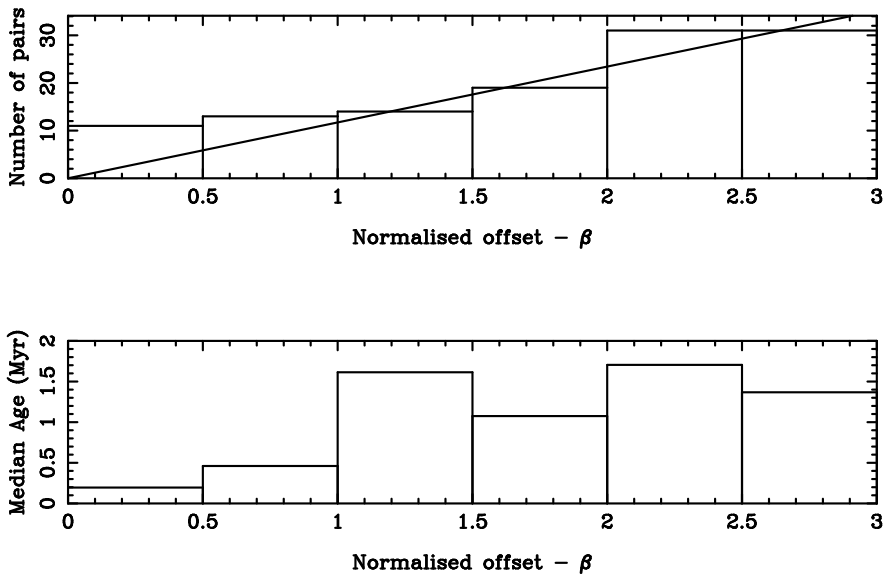}}
\end{picture}
\caption[]
{
Top panel: The {\it observed} distribution of $\beta$ for the sample
of pulsar and supernova remnants. When compared to the distribution
expected by chance shown by the solid line, this shows a clear excess
of pairs with $\beta \leq 1$. The lower panel shows the median
characteristic age of the pulsars in the pairs as a function of
$\beta$ (see text).
}
\label{fig:reals}
\end{figure*}

\section{Pulsar--Supernova Remnant Pair Statistics}
\label{sec:stats}

In this section, we address the question: How many pulsar-supernova
remnant pairs (which we shall hereafter refer to simply as pairs) are
likely to occur by chance in the present sample? As mentioned
previously, this question has been tackled in some detail by Gaensler
\& Johnston (1995a\&b) who concluded that the majority of claimed
associations are likely to be chance alignments. The main motivation
for the present analysis is partly to approach the question from a
different direction than the modelling of Gaensler \& Johnston
(1995b). Our analysis has the main advantage that it is somewhat
simpler and more model-free than the method described by Gaensler \&
Johnston (1995b).

For the purposes of the analysis it is useful to characterise each
pair by the dimensionless parameter $\beta$ which is independent of
distance, defined as the ratio of the angular separation between the
pulsar and the remnant centroid to the angular radius of the
remnant. Thus pairs in which the pulsar is within the boundary of the
remnant occur if $\beta \leq 1$, whereas $\beta >1$ indicates that the
pulsar lies outside the remnant (see also Shull et al.~1989; Frail et
al.~1994; Gaensler \& Johnston 1995 a\&b).

We are interested in deriving the distribution of $\beta$ that occurs
by chance and comparing this directly with the observed
distribution. For completely unrelated sets of pulsars and supernova
remnants, the number of pairs occupying an annulus between $\beta$ and
$\beta + d \, \beta$ is proportional to $\beta$, regardless of the
relative densities of pulsars and supernova remnants over the plane of
the sky. To demonstrate this, we decoupled the respective pulsar and
supernova remnant samples by applying a systematic shift to the
Galactic longitude of each pulsar and then calculating the
distribution of $\beta$.  To improve the statistics, we performed this
procedure for shifts of $\pm 4$ and $\pm 8$ degrees in the Galactic
longitudes of the pulsars. The shift sizes were chosen so that they
are small compared to changes in the density of both types of objects
on the sky, whilst being larger than the angular size of any supernova
remnant in Green's (1996) catalogue. The shifted samples therefore
contain only pairs which are truly unrelated, allowing us to deduce
the expected distribution of chance remnants.  This distribution is
shown in Fig.~\ref{fig:diddle} and is in excellent agreement with the
theoretical prediction $dN \propto \beta \, d \beta$ shown by the
straight line fit through the origin. The slope of the best-fit
straight line shown in Fig.~\ref{fig:diddle} is $47\pm3$.

We are now in a position to apply this relationship, which is based on
a superposition of four longitude-shifted samples, to the observed
distribution of $\beta$ shown in Fig.~\ref{fig:reals}. The straight
line shown in this case thus has a slope which is one quarter the
value derived above. Comparing the observed distribution with this
straight line we see a clear excess of pairs with $\beta \leq 1$ in
the observed sample, whilst the distribution with $\beta > 1$ is in
good agreement with the theoretical prediction. The pair excess is
clearly due to the fact that many of these are genuine associations,
not chance line-of-sight alignments. From Fig.~\ref{fig:reals}, we
infer that $12\pm5$ pairs are likely to be genuinely associated.

The difference between the shifted and observed samples can also be
seen in the lower panels of Figs. \ref{fig:diddle} and
\ref{fig:reals}, where we have plotted the median characteristic age
of the pulsars as a function of $\beta$. The shifted samples show no
significant deviation from a flat distribution with respect to
$\beta$, and have a median age of $\gapp 1$ Myr.  The observed sample
however is clearly more youthful for $\beta \leq 1$ than for $\beta >
1$ for which the median characteristic age is again $\gapp$ 1 Myr.

From an inspection of the observed sample and the available
literature, we have compiled a list of the most likely associations in
Table \ref{tab:top8}. A critical appraisal of each of these and other
proposed associations can be found in the original references and see
also Gaensler \& Johnston (1995b) and Kaspi (1996). \nocite{kas96}
Note that this list is concerned only with Galactic associations and
therefore does not include PSR B0540$-$69 associated with the
supernova remnant in the LMC \nocite{shh84} (Seward et
al.~1984). Based on the above analysis and the available statistics,
which suggested a total of $12\pm5$ pairs, we expect anything from
zero to nine further pairs to represent genuine associations. Examples
of further candidates include recently discovered remnants around
PSRs~B1643$-$43 and B1706$-$44 (Frail et al.~1994), PSR B1930+22 and
G57.3$-$1.2 \nocite{rv88} (Routledge \& Vaneldik 1988) and PSR
J0538+2817 and G180.0$-$1.7 \nocite{acj+96} (Anderson et al.~1996).
In the absence of further information to validate these associations,
we list the 8 sources in Table \ref{tab:top8}.

\begin{table}
\begin{center}
\begin{tabular}{lrrll}
\hline
Pulsar   & Period &   Age & $\beta$ &  Supernova \\
Name     & ms     &   kyr &         &  Remnant   \\
\hline
B0531+21   &  33   &   1.3 &   0.10  & G184.6$-$5.8 \\
B0833$-$45 &  89   &  11.4 &   0.29  & G263.9$-$3.3 \\
B1338$-$62 & 193   &  12.1 &   0.39  & G308.8$-$0.1 \\
B1757$-$24 & 125   &  15.5 &$\sim$1  & G5.4$-$1.2   \\
           &       &       &         &              \\
B1853+01   & 267   &  20.3 &   0.51  & G34.7+0.1    \\
B1509$-$58 & 151   &   1.6 &   0.24  & G320.4$-$1.2 \\
B1951+32   &  40   & 107.3 &   0.14  & G69.0+2.7    \\
B2334+61   & 495   &  40.9 &   0.08  & G114.3+0.3   \\
\hline
\end{tabular}
\end{center}
\caption[]
{
A compilation of the most likely genuine pulsar--supernova remnant
pairs with $\beta \leq 1$. Our statistical analysis suggests that, at
most, a further nine pairs may be real (see text).
}
\label{tab:top8}
\end{table}

As discussed in Sect.~\ref{ssec:assoc}, both of the newly discovered
pulsars in this survey, PSRs {\PSRA} and {\PSRB}, have characteristic
ages much larger than that expected for their target remnants. If they
were to be genuinely associated with these remnants, then both the
pulsars would have had to have been born with periods similar to the
presently observed values, 0.5 and 0.3 seconds, in order to explain
the large characteristic ages. In addition, the respective observed
value of $\beta$ for these pulsars 0.7 and 0.8 does {\it not} place
them in the group of pairs most likely to be genuinely associated.
Based on the results of the above analysis, there is no strong
statistical requirement for either of these pulsars to be associated
with the remnants although they cannot be entirely ruled out.

Whether a significant number of radio pulsars were born with such long
periods is controversial. Several authors, notably Vivekanand \&
Narayan (1981), \nocite{vn81} Narayan (1987), \nocite{nar87} Narayan
\& Ostriker (1990) \nocite{no90} and Deshpande et al.~(1995)
\nocite{drs95} have found evidence for ``injection'' of pulsars into
the population with periods $\sim 500$ ms, however other authors (Lyne
et al.~1985; Stollman 1987; Lorimer et al.~1993) find no requirement
for it. \nocite{lmt85,sto87b,lbdh93} From an inspection of the pulse
periods listed in Table \ref{tab:top8}, with the possible exception of
PSR B2334+61, none of the pulsars listed in Table~\ref{tab:top8} are
likely to have had such long periods at birth. Indeed Kulkarni et
al.~(1993) \nocite{kpha93} argue that PSR B2334+61 is the energy
source to G114.3+0.3, which appears to be a Crab-like nebula. In this
case, the initial period of B2334+61 must have been $\lapp 100$
ms. Thus the simplest, and most likely conclusion, to be drawn from
the present sample of pulsar--supernova remnant pairs is that they
support the notion that all pulsars are born with initial spin periods
$\lapp 100$ ms.

Finally we wish to point out that, whilst our method is in principle
sensitive to pairs over a large range in $\beta$, the sample that we
have used for our analysis is far from homogeneous. Therefore any real
pairs with $\beta \gapp 1$ are most likely to be underestimated in the
present sample in comparison with those with $\beta \lapp 1$ since
many of the targeted searches for pulsars in remnants have
concentrated mainly close to the remnant centre. Indeed, together with
the searches by Gorham et al.~(1996) and Kaspi et al.~(1996), our
search represents the first major effort to search the entire
area, rather than just the centroid,
of many of the more extended supernova remnants. Further
searches with improved sensitivity to pulsars with $\beta \gapp 1$
will improve the situation.  New multibeam searches of the Galactic
plane, presently underway at Parkes and Jodrell Bank, should provide a
more homogeneous sample for further statistical studies.

\section{Conclusions}
\label{sec:conc}

We have conducted a sensitive search for pulsars in supernova
remnants. The search detected a total of three pulsars, two of which
were previously unknown. The new pulsars, {\PSRA} and {\PSRB}, were
found during searches of the supernova remnants {\SNRA} and {\SNRB}
respectively. The case for associations between the new pulsars and
their target remnants is presently unclear.  From an analysis of the
present sample of radio pulsars and supernova remnants, we reach two
main conclusions: (i) The number of real associations in the present
sample is at most $12\pm5$. (ii) These are most likely to occur for
pairs with $\beta \lapp 1$ for which we see a clear excess in the
observed distribution compared with that expected by chance.

Based on this analysis, neither of the two newly discovered pulsars
seem likely to be genuinely associated. In order to unambiguously
confirm/refute both these candidate associations, measurements of the
pulsar proper motions are required.  Such measurements would determine
in each case whether the pulsar velocity is of the correct magnitude
and direction to carry them to their presently position with respect
to the remnant centre. Interferometric measurements to determine the
proper motion of both pulsars using MERLIN, the {\bf M}ulti--{\bf
E}lement {\bf R}adio {\bf L}inked {\bf I}nterferometer network
operated by the University of Manchester have recently been initiated.

\acknowledgements 

It is a pleasure to thank Christine Jordan and Jon Bell for help with
the observations, as well as Nichi D'Amico, Luciano Nicastro and Dick
Manchester for making their search software available to us and Lee
Hannan for his assistance with the flux density determinations. We
would also like to thank the anonymous referee for constructive and
useful comments.  During this project, we made frequent use of Green's
(1996) on-line catalogue of supernova remnants as well as the NASA
Astrophysics Data System.  Both DRL and FC acknowledge financial
support from the European Commission in the form of {\it European
Pulsar Network} fellowships, funded under the HCM contract nr.~ERB
CHRX CT940622. FC gratefully acknowledges present support from the
European Commission through a Marie Curie fellowship under contract
nr.~ERB FMBI CT961700.


\begin{thebibliography}{}

\bibitem[Albinson {\rm et~al.}<1986>]{atsg86}
Albinson~J.~S., Tuffs~R.~J., Swinbank~E., Gull~S.~F., 1986, MNRAS 219, 427

\bibitem[Anderson {\rm et~al.}<1996>]{acj+96}
Anderson~S., Cadwell~B.~J., Jacoby~B.~A. et al.
1996, ApJ 468, L55

\bibitem[Berkhuijsen<1987>]{ber87}
Berkhuijsen~E.~M., 1987, A\&A 181, 398

\bibitem[Biggs \& Lyne<1996>]{bl96}
Biggs~J.~D., Lyne~A.~G., 1996, MNRAS 282, 691

\bibitem[Biggs<1990>]{big90b}
Biggs~J.~D., 1990, MNRAS 245, 514

\bibitem[Brazier {\rm et~al.}<1996>]{bkc+96}
Brazier~K. T.~S., Kanbach~G., Carraminana~A., Guichard~J., Merck~M., 1996,
  MNRAS 281, 1033

\bibitem[Caraveo<1993>]{car93}
Caraveo~P.~A., 1993, ApJ 415, L111

\bibitem[Clark \& Caswell<1976>]{cc76}
Clark~D.~H., Caswell~J.~L., 1976, MNRAS 174, 267

\bibitem[Clifton {\rm et~al.}<1992>]{clj+92}
Clifton~T.~R., Lyne~A.~G., Jones~A.~W., McKenna~J., Ashworth~M., 1992, MNRAS
  254, 177

\bibitem[Coe, Jones \& Lehto<1994>]{cjl94}
Coe~M.~J., Jones~L.~R., Lehto~H., 1994, MNRAS 270, 178

\bibitem[Deshpande, Ramachandran \& Srinivasan<1995>]{drs95}
Deshpande~A.~A., Ramachandran~R., Srinivasan~G., 1995, JA\&A 16, 69

\bibitem[Dewey {\rm et~al.}<1984>]{dss+84}
Dewey~R., Stokes~G., Segelstein~D., Taylor~J., Weisberg~J., 1984, in
  Reynolds~S., Stinebring~D., eds, Millisecond Pulsars.
\newblock NRAO : Green Bank, p.~234

\bibitem[Fahlman \& Gregory<1981>]{fg81}
Fahlman~G.~G., Gregory~P.~C., 1981, Nat 293, 202

\bibitem[Feldt \& Green<1993>]{fg93}
Feldt~C., Green~D.~A., 1993, A\&A 274, 421

\bibitem[Frail \& Moffett<1993>]{fm93}
Frail~D.~A., Moffett~D.~A., 1993, ApJ 408, 637

\bibitem[Frail, Goss \& Whiteoak<1994>]{fgw94}
Frail~D.~A., Goss~W.~M., Whiteoak~J. B.~Z., 1994, ApJ 437, 781

\bibitem[F\"urst, Reich \& Seiradakis<1993>]{frs93}
F\"urst~E., Reich~W., Seiradakis~J.~H., 1993, A\&A 276, 470

\bibitem[Gaensler \& Johnston<1995a>]{gj95a}
Gaensler~B.~M., Johnston~S., 1995a, MNRAS 275, 73P

\bibitem[Gaensler \& Johnston<1995b>]{gj95c}
Gaensler~B.~M., Johnston~S., 1995b, MNRAS 277, 1243

\bibitem[Gold<1968>]{gol68}
Gold~T., 1968, Nat 218, 731

\bibitem[Gorham {\rm et~al.}<1996>]{gra+96}
Gorham~P.~W., Ray~P.~S., Anderson~S.~B., Kulkarni~S.~R., Prince~T.~A., 1996,
  ApJ 458, 257

\bibitem[Green<1989>]{gre89}
Green~D.~A., 1989, MNRAS 238, 737

\bibitem[Green<1991>]{gre91}
Green~D.~A., 1991, PASP 103, 209

\bibitem[Green<1996>]{gre96b}
Green~D.~A.
\newblock 1996.
\newblock Mullard {R}adio {A}stronomy {O}bservatory, {C}ambridge, {U}nited
  {K}ingdom (available on the {W}orld--{W}ide--{W}eb at {\it
  http://www.mras.cam.ac.uk/surveys/snrs/})

\bibitem[Green \& Gull<1989>]{gg89}
Green~D.~A., Gull~S.~F., 1989, MNRAS 237, 555

\bibitem[Gunn \& Ostriker<1970>]{go70}
Gunn~J.~E., Ostriker~J.~P., 1970, ApJ 160, 979

\bibitem[Hailey \& Craig<1994>]{hc94}
Hailey~C.~J., Craig~W.~W., 1994, ApJ 434, 635

\bibitem[Haslam {\rm et~al.}<1982>]{hssw82}
Haslam~C. G.~T., Salter~C.~J., Stoffel~H., Wilson~W.~E., 1982, A\&AS 47, 1

\bibitem[Helfand, Becker \& White<1995>]{hbw95}
Helfand~D.~J., Becker~R.~H., White~R.~L., 1995, ApJ 453, 741

\bibitem[Johnston {\rm et~al.}<1992>]{jlm+92}
Johnston~S., Lyne~A.~G., Manchester~R.~N. et al.
1992, MNRAS 255, 401

\bibitem[Kaspi<1996>]{kas96}
Kaspi~V.~M., 1996, in Bailes~M., Johnston~S., Walker~M., eds, Pulsars: Problems
  and Progress, {IAU} Colloqium 160.
\newblock Astronomical Society of the Pacific, p.~375

\bibitem[Kaspi {\rm et~al.}<1996>]{kmj+96}
Kaspi~V.~M., Manchester~R.~N., Johnston~S., Lyne~A.~G., D'Amico~N., 1996, AJ
  111, 2028

\bibitem[Kulkarni {\rm et~al.}<1993>]{kpha93}
Kulkarni~S.~R., Predehl~P., Hasinger~G., Aschenbach~B., 1993, Nat 362, 135

\bibitem[Landecker {\rm et~al.}<1989>]{lprv89}
Landecker~T.~L., Pineault~S., Routledge~D., Vaneldik~J.~F., 1989, MNRAS 237,
  277

\bibitem[Landecker, Roger \& Higgs<1980>]{lrh80}
Landecker~T.~L., Roger~R.~S., Higgs~L.~A., A\&AS 1980, 39, 133

\bibitem[Large, Vaughan \& Mills<1968>]{lvm68}
Large~M.~I., Vaughan~A.~E., Mills~B.~Y., 1968, Nat 220, 340

\bibitem[Lawson {\rm et~al.}<1987>]{lmop87}
Lawson~K.~D., Mayer~C.~J., Osborne~J.~L., Parkinson~M.~L., 1987, MNRAS 225,
  307

\bibitem[Leahy {\rm et~al.}<1985>]{lvln85}
Leahy~D.~A., Venkatesan~D., Long~K.~S., Naranan~S., 1985, ApJ 294, 183

\bibitem[Lorimer {\rm et~al.}<1993>]{lbdh93}
Lorimer~D.~R., Bailes~M., Dewey~R.~J., Harrison~P.~A., 1993, MNRAS 263, 403

\bibitem[Lorimer {\rm et~al.}<1995>]{lylg95}
Lorimer~D.~R., Yates~J.~A., Lyne~A.~G., Gould~D.~M., 1995, MNRAS 273, 411

\bibitem[Lozinskaya, Sitnik \& Pravdikova<1993>]{lsp93}
Lozinskaya~T.~A., Sitnik~T.~G., Pravdikova~V.~V., 1993, AZ 70, 469

\bibitem[Lyne<1988>]{lyn88}
Lyne~A.~G., 1988, in Schutz~B., ed, Gravitational Wave Data Analysis, ({NATO
  ASI Series}).
\newblock Reidel, Dordrecht, p.~95

\bibitem[Lyne \& Lorimer<1994>]{ll94}
Lyne~A.~G., Lorimer~D.~R., 1994, Nat 369, 127

\bibitem[Lyne \& Manchester<1988>]{lm88}
Lyne~A.~G., Manchester~R.~N., 1988, MNRAS 234, 477

\bibitem[Lyne, Anderson \& Salter<1982>]{las82}
Lyne~A.~G., Anderson~B., Salter~M.~J., 1982, MNRAS 201, 503

\bibitem[Lyne, Manchester \& Taylor<1985>]{lmt85}
Lyne~A.~G., Manchester~R.~N., Taylor~J.~H., 1985, MNRAS 213, 613

\bibitem[Manchester \& Taylor<1977>]{mt77}
Manchester~R.~N., Taylor~J.~H., 1977, Pulsars.
\newblock Freeman, San Francisco

\bibitem[Manchester {\rm et~al.}<1996>]{mld+96} Manchester~R.~N.,
Lyne~A.~G., D'Amico~N. et al.
1996, MNRAS 279, 1235

\bibitem[Narayan<1987>]{nar87}
Narayan~R., 1987, ApJ 319, 162

\bibitem[Narayan \& Ostriker<1990>]{no90}
Narayan~R., Ostriker~J.~P., 1990, ApJ 352, 222

\bibitem[Narayan \& Schaudt<1988>]{ns88}
Narayan~R., Schaudt~K.~J., 1988, ApJ 325, L43

\bibitem[Pacini<1968>]{pac68}
Pacini~F., 1968, Nat 219, 145

\bibitem[Pineault {\rm et~al.}<1993>]{plmb93}
Pineault~S., Landecker~T.~L., Madore~B., Gaumont-Guay~S., 1993, AJ 105, 1060

\bibitem[Reich \& Braunsfurth<1981>]{rb81b}
Reich~W., Braunsfurth~E., 1981, A\&A 99, 17

\bibitem[Reich, Fuerst \& Arnal<1992>]{rfa92}
Reich~W., Fuerst~E., Arnal~E.~M., 1992, A\&A 256, 214

\bibitem[Roberts {\rm et~al.}<1993>]{rgk+93}
Roberts~D.~A., Goss~W.~M., Kalberla~P.~M., Herbstmeier~U., Schwarz~U.~J., 1993,
  A\&A 274, 427

\bibitem[Routledge \& Vaneldik<1988>]{rv88}
Routledge~D., Vaneldik~J.~F., 1988, ApJ 326, 751

\bibitem[Routledge, Landecker \& Vaneldik<1986>]{rlv86}
Routledge~D., Landecker~T.~L., Vaneldik~J.~F., 1986, A\&A 221, 809

\bibitem[Routledge {\rm et~al.}<1991>]{rdlv91}
Routledge~D., Dewdney~P.~E., Landecker~T.~L., Vaneldik~J.~F., 1991, A\&A 247,
  529

\bibitem[Seward, Harnden \& Helfand<1984>]{shh84}
Seward~F.~D., Harnden~F.~R., Helfand~D.~J., 1984, ApJ 287, L19

\bibitem[Shemar \& Lyne<1996>]{sl96}
Shemar~S.~L., Lyne~A.~G., 1996, MNRAS 282, 677

\bibitem[Shull, Fesen \& Saken<1989>]{sfs89}
Shull~J.~M., Fesen~R.~A., Saken~J.~M., 1989, ApJ 346, 860

\bibitem[Staelin \& Reifenstein<1968>]{sr68}
Staelin~D.~H., Reifenstein~{III}~E.~C., 1968, Sci 162, 1481

\bibitem[Stollman<1987>]{sto87b}
Stollman~G.~M., 1987, A\&A 178, 143

\bibitem[Tatematsu {\rm et~al.}<1990>]{tflr90}
Tatematsu~K., Fukui~Y., Landecker~T.~L., Roger~R.~S., 1990, A\&A 237, 189

\bibitem[Tauris \& Manchester<1997>]{tm97}
Tauris~T.~M., Manchester~R.~N., 1997, MNRAS in press

\bibitem[Taylor \& Cordes<1993>]{tc93}
Taylor~J.~H., Cordes~J.~M., 1993, ApJ 411, 674

\bibitem[van~den Bergh<1971>]{van71}
van~den Bergh~S., 1971, ApJ 165, 457

\bibitem[Vivekanand \& Narayan<1981>]{vn81}
Vivekanand~M., Narayan~R., 1981, JA\&A 2, 315

\end{thebibliography}
\end{document}